\documentclass{svmult}
\usepackage{epsfig,graphicx} % figures
\begin{document}
\title*{Solar physics at the Kodaikanal Observatory: A Historical Perspective}
\author{S. S. Hasan, D.C.V. Mallik, S. P. Bagare \& S. P. Rajaguru}
%\authorindex{Priest, E.}
\institute{Indian Institute of Astrophysics, Bangalore, India}
%\institute{School of Mathematics and Statistics, 
%     St Andrews University, Scotland} 
\maketitle

%\vspace*{-20ex}
\section{Background}
The Kodaikanal Observatory traces its origins to the East India Company which 
started an observatory in Madras ``for promoting the knowledge of astronomy, geography and navigation
in India''.   Observations began in 1787 at the initiative of William Petrie, an officer of the Company, with the use of  two 3-in achromatic telescopes, two astronomical clocks with compound penduumns and a transit instrument.  By the early 19th century the Madras Observatory had already established a reputation as a leading astronomical centre 
devoted to work on the fundamental positions of stars, and 
a principal source of stellar positions for most of the southern hemisphere stars.  John
Goldingham (1796 - 1805, 1812 - 1830), T. G. Taylor (1830 - 1848), W. S. Jacob (1849 - 1858)
and Norman R. Pogson (1861 - 1891) were successive Government Astronomers who
led the activities in Madras. Scientific highlights of the work included a catalogue of 11,000 southern stars  produced by the Madras Observatory in 1844 under Taylor's direction using the new 5-ft transit instrument. 

The observatory had recently acquired a transit circle by Troughton and Simms which was mounted and ready for use in 1862. Norman Pogson, a well known astronomer whose name is associated with the modern definition of the magnitude scale and who had considerable experience with transit instruments in England, put this instrument to good use. With the help of his Indian assistants, Pogson measured
accurate positions of about 50,000 stars from 1861 until his death in 1891. During this period two total eclipses and one annular eclipse of the Sun were
visible from India. Pogson led teams  to all three of them. The first one of these, a total eclipse on August 18, 1868, created an enormous interest amongst European astronomers and preparations for its observation were made in England and France for many months preceding the event. Teams of professional astronomers from both countries arrived in India and established their camps at Guntoor, on the central line of the eclipse. The Madras Observatory astronomers had their camp at
Masulipatam and Vunpurthy further east. This eclipse is of great historical significance as it
was the first time when spectroscopes were used during an eclipse event. A new line close
to the D$_2$ line of sodium and to the left of it was seen in the spectrum of the chromosphere.
PogsonÕs hand drawing of the line is shown in the figure.

\begin{figure}  
\begin{center}
\includegraphics[width=0.6\textwidth]{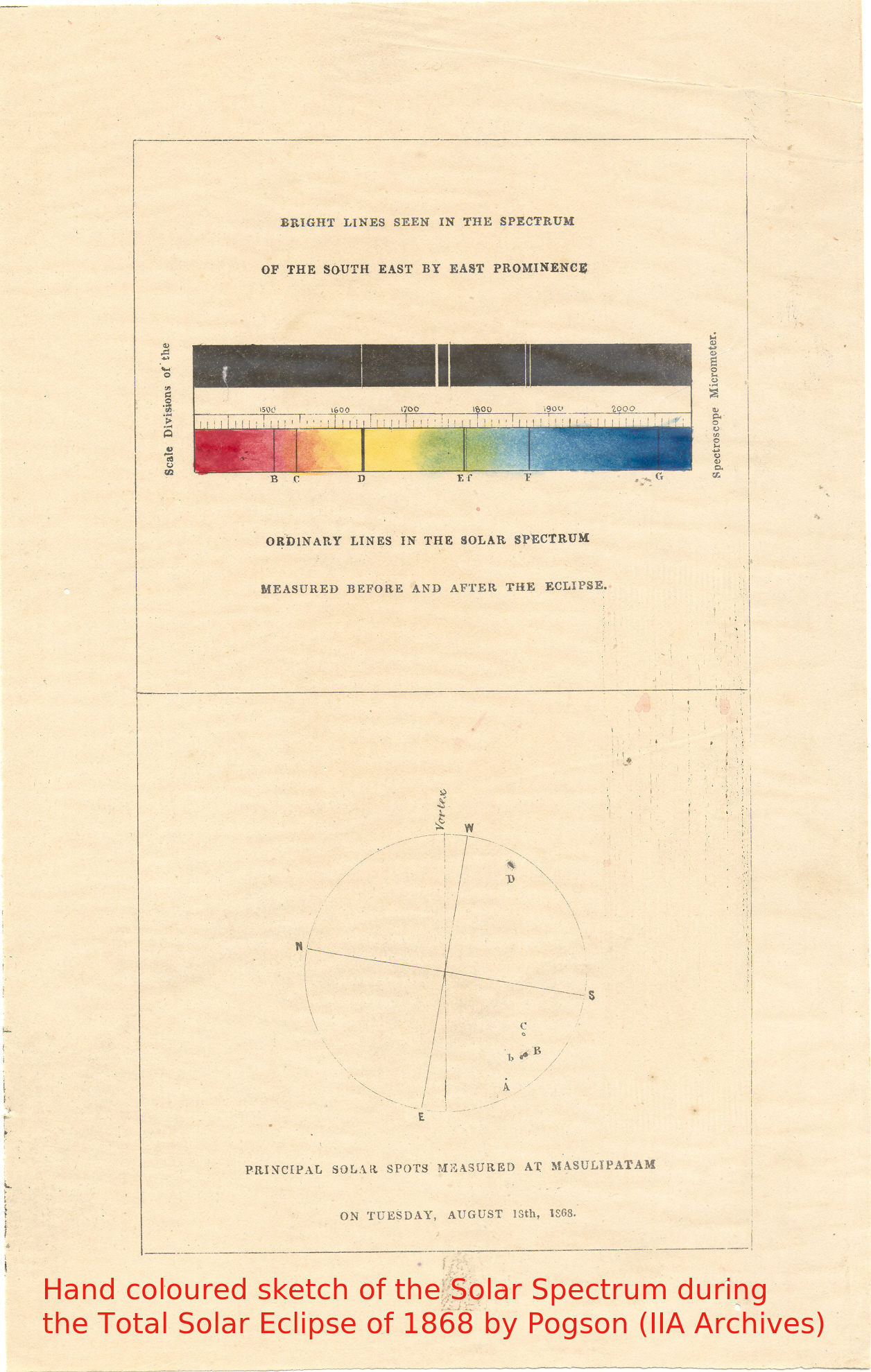}
\caption[]{} 
\end{center}
\end{figure}

The discrepancy in the wavelength of this line with the sodium line was confirmed by Norman
Lockyer who could not ascribe it to any known terrestrial element. This observation, in fact, marks the discovery of helium. During the same eclipse, observations of the hydrogen Balmer lines in the spectra of the prominences established their gaseous nature.

Historically, the eclipse of 1868  is an important landmark associated with the birth of solar physics in India. Janssen and Lockyer
made effective use of the spectrograph to show that ``red prominences could at any time be
examined, without waiting for an eclipse at all''. During the annular eclipse of June 6, 1872,
which was visible from Madras, Pogson found the bright chromospheric spectrum flash out
for a short duration on the formation and again at the breaking up of the annulus. This is the
first observation on record of viewing the flash spectrum at an annular eclipse.

In May 1882, Pogson  proposed a 20-in telescope to augment the needs of his observatory. The proposal had received active support both in India and Britain and the search for a suitable location in the southern highlands of India was authorised. The task was entrusted to Michie Smith, Professor of Physics in the Christian College, Madras, who had arrived in India in 1877. Michie Smith undertook a survey of the
Palni and Nilgiri Hills in 1883 to 1885, his observations covering the twin requirements of
transparency and steadiness of image during both day and night. However, there was a delay in the project being cleared by the Astronomer Royal in England since he felt ``saddling
Pogson with additional work connected with the new large equatorial, when he has accumulated
large arrears in observations, would not be desirable.''

In 1879, a Committee was appointed by the British Government, consisting of some leading
astronomers in the country, to consider and advise on the methods of carrying on observations
in solar physics. One of the tasks of the Committee, which came to be known as the Solar
Physics Committee (hereafter, SPC), was to reduce the solar photographs taken daily since
1878 in Dehra Dun with a photoheliograph. The British Government in India had supported
the work in the belief that a study of the Sun would help in the prediction of the monsoons,
their success and failure, the latter often leading to famines that caused such a havoc. SPC
also suggested to the Government of India ``that photographs of the Sun should be taken
frequently in order that India might assist towards securing a permanent record of the number
and magnitude of the sunspots and other changes in the solar surface". At a later date they
added that special spectroscopic observations should also be taken in India, in order to collect
evidence which will probably throw light on the constitution of the Sun.
In 1885, the Royal Society, London, constituted the Indian Observatories Committee comprised
of the Astronomer Royal, and a few Fellows of the Royal and Royal Astronomical Societies.

The Committee was entrusted with the task of coordinating the work of Madras and Bombay
Observatories and of advising the Secretary of State for India pertaining to the administration
of these observatories. The suggestions of SPC regarding regular observations of the Sun and
the thinking of the Indian Observatories Committee on the future orientation of activities of
the Madras Observatory converged to pave the way for the creation of a new observatory
in the hills of South India. After the death of Norman Pogson on June 23, 1891, a series of
initiatives were taken that eventually led to the establishment of the Solar Physics Observatory
in Kodaikanal, a change in the administrative control of the observatories and the relegation
of the Madras Observatory to a secondary status. It took about nine years to effect this
transition.

\section{Birth of the Kodaikanal Observatory}
In October 1891, John Eliot, the Meteorological Reporter to the Government of India wrote to
the Government of Madras about a rethinking of the status of the Madras Observatory by
the Secretary of State, Government of India, on the advice of the Astronomer Royal. They felt
it was desirable that, in place of the Madras Observatory, a large Astronomical Observatory
should be maintained in India, the work at which should, as defined by the Astronomer Royal
in a letter, dated April 14, 1883, consist of (a) astronomical work proper, including the
determination of accurate positions of the Sun, certain fundamental stars, the moon and
planets, and (b) astronomical physics, more especially, daily observation of the solar
prominences and other solar phenomena with the spectroscope, and daily photography of the
Sun. Eliot suggested that the work on the Sun as recommended by SPC should get high
priority and there was a need to relocate the observatory to a suitable hill-station, where the
climate was nicer and more conducive to work. He further suggested that changes and
additions of the European staff would be required for the observatory to carry out both solar
physics and other astronomical work and changes in the control and inspection of the
observatory would be necessary to ensure its working in the way desired by the Secretary
of State, Governmentt of India and the Observatories Committee in England.

A serious effort was then made to find a suitable location for an observatory. The regions
chosen for further exploration were Kotagiri in the Nilgiris and Kodaikanal in the Palnis. On
PogsonÕs death, The Government of
Madras directed Michie Smith to conduct the site survey at Kodaikanal and Kotagiri.
In August 1892, Michie Smith submitted to the Government of Madras a detailed report on
the relative merits of Kodaikanal, Kotagiri and Madras as sites for an astronomical observatory.
He had based the report on extensive observations of (i) the Sun with respect to faculae, spots and their spectra, (ii) of Venus and Saturn, (iii) star clusters and nebulae, (iv) double stars, and (v)
measurements of apparent magnitudes of stars and (vi) photographs of star trails. The steadiness
of the image of the Sun was also studied. On all counts, Kodaikanal appeared to be the best
of the three sites. It also had the advantage of altitude, the spot chosen there by Smith was
more than a 1000 ft higher than the highest spot in Kotagiri.

The report was considered by the SPC in July 1892 and by the Indian Observatories Committee
under the Chairmanship of Lord Kelvin in October of the same year. SPC suggested that if
funds were not available for equipping a complete new observatory with two branches, the
furtherance of solar physics in India would be the greater benefit to science. Continuity of
the photographic registration of the sunspots in India had to be maintained as it was only
with the Indian cooperation the then nearly complete record had been secured. Both the
committees endorsed the appointment of Michie Smith on a more permanent basis to superintend
the work at the new observatory. It was also said that the magnetic and meteorological
observations could continue at Madras, which would become a branch observatory with a
superintendent and a small staff. It was asserted that with these changes, the Southern India
Astronomical Observatory would become the only Government Astronomical and Solar Physics
Observatory in India. It would therefore be a national and not a provincial institution. The
committees felt it was essential that the control of the new institution should go to the Imperial
Government.

In a gazette notification dated November 21, 1893, it was stated that Kodaikanal afforded the
most suitable site for the proposed Solar Physics Observatory. The Meteorological Reporter
suggested that the consideration of the question of the permanent site of the Astrononomical
Observatory be deferred for some five years. In 1894, Michie Smith's position as the Government
Astronomer became permanent.
With the official sanction and notification, actual planning of the observatory started. Michie
Smith was allowed additional duty leave to spend time in England, from May 17 to October
9, 1895, to enable him to discuss with the astronomers in Europe many issues pertaining to
the setting up of the observatory. On July 17, 1895, Michie Smith laid before the Indian
Observatories Committee his plans for the proposed Kodaikanal Observatory and the equipment.
The Committee strongly recommended the acquiring of a solar spectroscope. The other major
equipment, already collected in Madras from various sources, that had to be shifted to
Kodaikanal, consisted of a photoheliograph, a spectrograph, a 6-in Cooke Equatorial, a
transit telescope and chronograph, and a 6-in Lerebour and Secretan equatorial.
In October 1895, the foundation stone of the new observatory was laid by the Governor of
Madras. By the end of 1896 plans and estimates of the new establishment were completed and
sanctioned. In July 1897 Michie Smith visited Kodaikanal and laid out a North-South line for
the observatory building. The building work was slow to begin with but picked up later. When
the large spectrograph Ñ a polar siderostat with an 11-in mirror, a 6-in lens of 40-ft
focal length and a concave grating mounted on RowlandÕs plan, was received in 1897, it had
to be temporarily housed in the library building.

On January 22, 1898 there was another total solar eclipse at Sahdol, in the princely state of
Rewa. The Secretary of State invited the Astronomer Royal and Sir Norman Lockyer to visit
and also report on the various Indian observatories. The Astronomer Royal, W. H. M. Christie,
visited Madras early in February (1898) on his way home from his eclipse camp at Sahdol. 
Christie spent two days at the Observatory and carefully went into all questions concerning
the equipment and the work that was being carried on. He then visited Kodaikanal and
discussed the plans of the new buildings. Various changes which would improve the buildings
were suggested and were at once adopted. Sir Norman Lockyer visited Madras but not
Kodaikanal and expressed his dissenting views on the planned buildings in Kodaikanal. On
his return to England, Lockyer represented to the Secretary of State for India that the buildings
were too costly and too permanent, while the structures in his own South Kensington
observatory were temporary and cheaper, although his observatory was the most powerful
solar observatory in the world. The Secretary of State ordered stoppage of all work at
Kodaikanal until the matter was sorted out with the Astronomer Royal.
Whether Lockyer's protest was forwarded to the Indian Observatories Committee is not
known, but the result was a delay from June to October end, after which the buildings were
allowed to go on. Books and instruments were transferred from Madras to Kodaikanal and sent
up the ghaut in the dry weather before end of March 1899. About 1000 coolie loads reached
Kodaikanal. No damage was detected. Michie Smith took up residence in Kodaikanal by the end
February 1899 as it was necessary for him to be there to advise the engineer in charge of
arrival.

Officially, the observatory started work on April 1, 1899. After the work of construction was
largely completed by December 1901, additional measures were taken to protect the Observatory
from very strong winds. Progress was made in planting oak and pine trees and laying out
grounds, although it was known that many years would elapse before these would take their
full effect in modifying the strength of the winds to which the Observatory was exposed.
In 1902 a Calcium-K spectroheliograph was ordered from Cambridge Scientific Instrument
Company. It arrived in Kodaikanal in 1904. The spectroheliograph consisted of a 12-in triple
achromatic lens of 20-ft focal length fed by a Foucault siderostat incoroporating an 18-in
aperture plane silver-on-glass mirror, fitted with differential electric motors for fine adjustments
in R.A. and Dec. The design of the instrument was as specified by George Ellery Hale (the
discoveror of the spectroheliograph), in which the image of the Sun and the camera remain
stationary while the collimating lens, the camera lens, the prisims and slits are carried on a
rigid frame which moves at right angles to the optical axis. This was the instrument that John
Evershed put to effective use after his arrival in Kodaikanal in 1907.
John Evershed joined on January 21, 1907 as Chief Assistant to the Director. Michie Smith
went on combined privilege leave and furlough for nine months from April 1. Naturally,
Evershed took charge of the observatory as soon as he arrived. This was his first professional
appointment as an astronomer.

\section{John Evershed (1864-1956)}
John Evershed was born at Gomshall in Surrey on February 26, 1864. He is a descendant of
the ancient family of Evereshavedes in Surrey who can be traced back to before the Norman
conquest (1066 A.D.). They were yeoman farmers and EvershedÕs grandfather was the last of the branch to carry on an unbroken family tradition as yeoman farmers in Surrey
for over 600 years. Evershed's motherÕs side of the family was in Portsmouth where his
maternal great grandfather had a shop selling nautical almanacs and charts. He was well known
to Lord Nelson.

Evershed's  interest in astronomy began rather early. When he was only 11, there was a partial
eclipse of the Sun, visible from Surrey, and boy Evershed ran almost all the way from Gomshall
to Shere to watch the eclipse using a telescope belonging to a doctor. Even before that when he
was six, his imagination was spurred by a picture of German shells falling in the streets of Paris
in 1868, during which Janssen, the French astronomer, escaped by balloon from the besieged
city to watch the total eclipse of the Sun. It was during this eclipse Janssen had discovered
the H$_\alpha$\  line in a solar prominence and soon after obtained pictures of the same outside
of an eclipse. This was the beginning of the daily spectroscopic observations of prominences
without waiting for an eclipse, an activity to which Evershed had a lot to contribute when he
grew up.

Evershed was, until his appointment in Kodaikanal, an accomplished amateur astronomer.  In the
1890s, when Evershed was already studying solar radiation, he
was induced by some great spectroscopists, including Pringsheim and Paschen, to perform
experiments on heated gases and study their radiation pattern. As he says ``I was able to show
that coloured vapours of iodine and bromine heated to the temperature of red heat glowed
with a continuous spectrum, at the same time giving a discontinuous absorption spectrum by
transmitted light. ... I was able to show that vaporised sodium could be made to emit its
characteristic D radiation by heat alone under conditions where there could be no action other
than heat.''

In 1890, Evershed set up a private observatory in Kenley, Surrey, and with the spectroscope,
he had designed, and his 3-in telescope he started a long series of observations of
prominences and recorded their distribution in heliographic latitude and their variation with
the cycle of sunspots. Between 1890 and 1905, Evershed recorded some 13,458 prominences.
In 1890, he became a founder member of the British Astronomical Association and he was the
Director of the Section on Solar Spectroscopy between 1893 and 1899.

The 1890s were exciting years for solar physics. In the spring of 1891, George Ellery Hale
invented the spectroheliograph. Like Evershed he had set up his own Kenwood Physical
Observatory in the backyard of his house in Chicago, where he had a self-made 12-in
telescope. Using the spectroheliograph and this telescope, he was soon able to see the
reversals in the Ca II H and K lines. Evershed had his own way of photographing prominences
in the H$_\beta$\  line of hydrogen. With this technique of monochromatic photography, he was getting
solar images showing the brilliant flocculi around sunspots. When he learnt about HaleÕs new instrument, he abandoned the idea of working on the hydrogen line and transferred
his attention to the Ca II K line. 

Evershed got to know Arthur Cowper Ranyard, a barrister by profession with a deep
interest in astronomy, who was the editor of Knowledge. Ranyard, who knew
Hale personally, introduced him to Evershed when the former visited England in 1894.
This was the beginning of a long friendship between the two pioneer solar astronomers of
the time, which lasted till Hale's death in 1938.
Ranyard died in 1894 and left Evershed with his 18-in reflector and a small spectroheliograph.
Evershed found problems with Ranyard's instruments which produced curved spectral lines and
gave distorted images of the Sun. He put the 18-in telescope at Kenley, and designed a
spectroheliograph on his own using large direct vision prisms which gave straight spectral
lines and undistorted solar images.

Evershed's immediate superior, in the company he was working, was interested in science and Evershed was
granted leave to join solar eclipse expeditions. His first visit to India was in connection with
the eclipse of January 22, 1898. He was in the company of W. H. M. Christie, the Astronomer
Royal, H. H. Turner, the Savilian Professor of Astronomy at Oxford, Sir Norman Lockyer
among others. He obtained excellent flash spectra and caught the emission continuum at the
head of the Balmer series extending to the ultraviolet end of the plate. He realised this was
the counterpart of the continuous absorption spectrum seen by William Huggins in stars with
strong hydrogen absorption lines. At a later solar eclipse in Algeria in 1900, he repeated the
same experiment and again saw the continuum emission. His results definitively proved that
the flash spectrum and the Fraunhofer spectrum were of the same origin, the flash spectrum
representing the higher and more diffused portion of the gases which by their absorption gave
the Frauhofer dark line spectrum. At all eclipses, Evershed carried his own home-made
instruments -- prismatic spectroscopes with a long focal length camera, where only the optical
parts were procured from specialised manufacturers. These prismatic spectroscopes consisting
of two or more prisms are often referred to as Evershed spectroscopes.

The results of EvershedÕs eclipse expeditions attracted the attention of the great stellar
spectroscopist Sir William Huggins who was then the President of the Royal Society. Evershed
and Huggins had an interesting correspondence in this matter. Later it was Huggins who
recommended the appointment of Evershed to the position of the chief assistant to Michie
Smith at the Kodaikanal Observatory .

In the original recommendation for setting up an Imperial Astronomical Observatory in the hills
of Southern India, the Government had indicated that the observatory would be headed by
an Astronomer to be appointed by the Indian Observatories Committee in England while the
Solar Physics Section would be headed by a Superintendent, either appointed by the Solar
Physics Committee and the Observatories Committee, or  Michie Smith, the then Government
Astronomer, could be appointed to the position. Smith was deemed to have the requisite
qualification and experience to run the Solar Observatory, while his suitability to head the
Imperial Observatory was somewhat in doubt. It was agreed that the staff of the new observatory
should consist of two Europeans and a group of Indian assistants. When the idea of establishing
an Imperial Observatory was temporarily abandoned, and it was decided that the Solar Physics
Observatory was the one to be immediately established in Kodaikanal, Michie Smith became
the natural choice for its directorship. The European assistant to him was not immediately
appointed.

The principal thrust of the work at Kodaikanal was to be solar spectroscopy. There were three
existing spectroscopes none of which appeared suitable for obtaining spectra of prominences
and sunspots with good efficiency. In 1903, it was decided to build a spectroscope and one
of the existing ones (a large six-prism instrument with a collimator and a telescope of 17-in
focus) was dismantled and two lenses from it were mounted with an existing diffraction grating
to rig up an instrument which was then used with the 6-in Lerebour and Secretan equatorial
to obtain spectra of the solar features. A three-prism Evershed spectroscope was ordered from
Hilger Co. during the same year and this went into operation in November 1904. It gave
excellent spectra of sunspots and prominences. Also in 1904, the Cambridge spectroheliograph
arrived in Kodaikanal. Naturally there was a great need of having a person who would direct
the spectroscopic work. Michie Smith visited Hale at Mt. Wilson in early 1904 and discussed
in detail the kind of work Kodaikanal should undertake. Hale was also keen on developing
an international network on solar research and he was particular that the work with
spectroheliographs, at Catania and Yerkes Observatories, his own at Mt. Wilson, yet to be
commissioned, and the newly acquired one at Kodaikanal, was done in a coordinated fashion.
He was already in correspondence with Evershed and was coaxing him to join this proposed
network, since Evershed had his own spectroheliograph and was producing Ca II K
spectroheliograms on a regular basis. Hale narrated in detail his conversations with Michie
Smith in a letter to Evershed on February 2, 1904. Evershed was quick to respond saying that
he felt a cooperation in spectroscopic and spectrographic methods of observation rather than
the other older methods of photographic work would indeed be of great help to the community.
He called himself Ôan irresponsible amateurÕ who was not afraid to be part of any professional
network.

\subsection{Evershed in Kodaikanal}
In 1904, the Government of India sanctioned the appointment of an European Assistant to the
Director, Michie Smith. John Evershed was clearly the most eminent choice for the position.
According to Evershed, Sir William Huggins greatly influenced the decision of the Government
of India in making Evershed the offer, which also had the support of Gilbert Walker, the
Director-General of Observatories in India. In a letter to Hale dated December 3, 1904, Evershed
said that he had been offered an appointment in Kodaikanal and it would take him six months
to decide. He was a bit depressed at the prospect of having to close down his private
observatory in Kenley. He probably had in mind a personal matter, that of his marriage, before
he could definitely take a final decision. He was preparing for the journey to India in 1906.
On the advice of Professor Turner of Oxford, he decided to go to India not by the regular
route, but by the longer Trans-Atlantic/Trans-Pacific one via America and Japan. Turner gave
Evershed introductions to the leading American astronomers of the day, Pickering in Harvard,
Frost at Yerkes and Barnard and Campbell at Lick. Turner also arranged for his visits to
Harvard and Princeton, and to these observatories and arranged for the India Office to
sanction these visits. Evershed's main aim, of course, was to spend a month with Hale at Mt.
Wilson.

At the eclipse expedition of August 9, 1896 in Norway, Evershed had first met Mary Ackworth
Orr, a young and sprightly amateur astronomer and a member of the British Astronomical
Association. Mary was also interested in Dante Alighieri and was to write a book titled ÔDante
and the Early AstronomersÕ in 1914. In 1906 September John and Mary got married. In a letter
to Hale on June 28, 1906, Evershed had told him about his plan to visit America on his way
to India. He also wrote that before leaving England he was to be married to ÔMary A OrrÕ who
would accompany him to America. On September 22, 1906, the Eversheds left England by the
Anchor Liner Columbia reaching New York end of September. The couple travelled to Williams
Bay, Wisconsin, where the Yerkes Observatory is situated. Edwin Frost, the Director hosted
them for a week. They arrived in Pasadena on October 18 planning to spend several weeks
with Hale. However, owing to some mis-communication Hale was away at the time and was
not returning to Mt. Wilson until the end of November. But the couple had a wonderful and
scientifically fruitful time organised by Hale's main solar collaborator Ellerman. Evershed was
deeply impressed by the instruments, the method of work and Hale's organisation of the
observatory. On January 21, 1907 the Eversheds reached Kodaikanal.

\begin{figure}  
\begin{center}
\includegraphics[width=\textwidth]{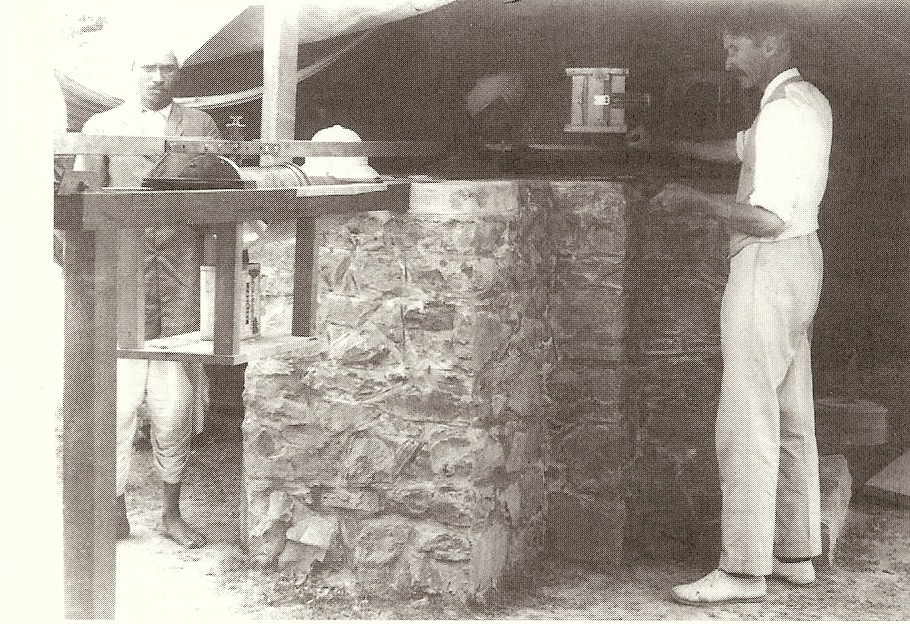}
\caption[]{John Evershed working with the spectroheliograph at the Kodaikanal Observatory (from IIA archives)} 
\end{center}
\end{figure}

Evershed immediately got involved in solar observations as soon as he arrived in Kodaikanal.The Cambridge
spectroheliograph had been put into operation in 1905. The building where it was housed was
faulty with a leaky roof and a new roof had to be constructed. There was also some problem
with the setting of the second slit at the correct wavelength. So, although spectroheliograms
were obtained in 1905 and 1906, the instrument did not perform to its full potential. It was left
to Evershed to bring the instrument to its fine working order. By then a new moving roof was
in place in the spectroheliograph building with an addition of ruberoid sheets to prevent leaks.
A new collimator slit was fitted and Evershed solved the problem with the camera slit, giving
it greater stability and incoroprating a new device to facilitate the setting of the slit at any
desired wavelength. He also started working on an auxiliary spectroheliograph to obtain
pictures of the Sun in H-alpha.

When Comet Daniel showed up in the sky, Evershed used his eclipse prisms (he had brought
them with him to India) and made a prismatic camera fitted to the 6-in Cooke equatorial to
obtain spectra of the comet. He identified the CN bands in both the nucleus and the tail of
the comet. When Halley's Comet appeared in 1910, Kodaikanal geared up to observe the object
both photographically and spectroscopically. Between April 19 and May 16, 1910 when the
comet was most amenable to observations in the early dawn, Evershed was able to obtain
spectra which showed the CN bands and the C2 Swan bands in the head, while the tail showed
bands due to CO. He was ably assisted in the observations by his wife Mary.

\begin{figure}  
\begin{center}
\includegraphics[width=0.5\textwidth]{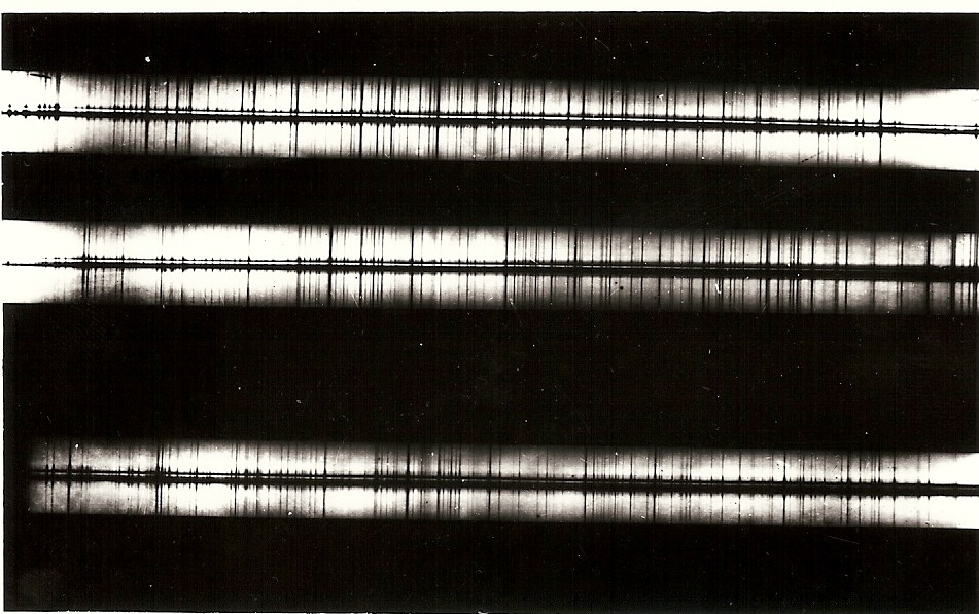}
\includegraphics[width=0.8\textwidth]{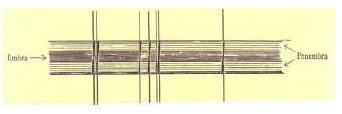}
\caption[]{Top: Solar spectra of a sunspot region recorded by Evershed on January 5, 1909 at Kodaikanal. Bottom: Line sketch of the spectrum showing the shift of the absorption line penumbra around the sunspot (from IIA archives)} 
\end{center}
\end{figure}

\subsection{Evershed Effect}
However, Evershed's chief work in the initial years in Kodaikanal was on the spectra of
sunspots. He used the high dispersion spectrographs at the observatory to systematically
study the spectra of spots whenever the weather was favourable. He noted that the sunpsot
spectra were constant in character and consisted mainly of intensified Fraunhofer lines. The
general absorption was continuous and was similar to the continuous absorption seen in the
limb of the Sun. The pressure in the spot regions seemed to be slightly less than in the
reversing layer of the photosphere. In 1909, Evershed made another Ôvery powerful
spectrographÕ, one he called spectrograph III. This had a parabolic silver-on-glass mirror
which formed the solar image on the slit plate and he used an excellent 6-in plane grating
by Michelson, which the observatory had, as the dispersing element. He employed this
instrument to solely concentrate on the spectra of sunspots.

In the early morning of January 5, 1909, on a day when the atmosphere was exceptionally
steady and the sky transparency was excellent after some heavy thunderstorm activity, Evershed
found two large spots and obtained their spectra. As he says, ``the spectra revealed a curious
twist in the lines crossing the spots which I at once thought must indicate a rotation of gases,
as required by Hale's recent discovery of strong magnetic field in spots, but it soon turned
out from spectra taken with the slit placed at different angles across a spot that the displacement
of the lines, if attributed to motion, could only be due to a radial accelerating motion outward
from the centre to the umbra. Later photographs of the Calcium H and K lines and the H$_\alpha$
line revealed a contrary or inward motion at the higher levels represented by these lines.'' This
was the first time line displacements in the penumbral region were seen indicating an outward
radial flow of gases in the spots. None of the previous spectroscopic studies had revealed
the dynamics of the flow of gases in the spots. Two days later, Evershed obtained more
spectra which confirmed the discovery.

Earlier studies had concentrated on spots near the central meridian of the Sun. Evershed was
the first to observe spots at various positions, up to 50 degrees, on either side of the central
meridian. He found that the line displacements were more pronounced in the penumbral
regions of the spots which are closer to the limb of the Sun. This established that the flow
of the gases was radially outward parallel to the surface of the Sun. In a letter to Hale on
February 1, 1909, Evershed described his remarkable discovery and said ``I have found that
there is a Doppler shift of all the lines in all spot spectra. I made a special study of it having
obtained more than one hundred and fifty spectra representing seven northern spots and four
southern.'' The complete results were published in the Kodaikanal Observatory Bulletins of
1909 and the inverse flow seen in the strong lines in a paper in MNRAS in 1910.

Hale responded soon congratulating Evershed for the discovery and stated that because of
the pressure of their magnetic field work, they were unable to investigate the dynamics of
gases in sunspots.

\subsection{Spectral lines and diagnostics}
The general problem of an unexplained redshift of spectral lines seen at the limb of the Sun
with respect to the same lines seen from the centre had occupied the thoughts of all major
observers at the time. This was first brought out by Halm at the Cape Observatory in 1907.
The Mt. Wilson astronomers, who investigated the phenomenon in great detail, came to the
conclusion that the shift was due to the differences in pressure in the two regions Ñ disk
centre and the limb. Evershed too was trying to understand the nature and cause of this shift.
His observations however indicated complexity. His work further showed that the pressure
interpretation was incorrect, as some of the lines supposed to be most affected by pressure
were actually shifted to the violet. Evershed had developed very accurate methods of measuring
extremely small line shifts and much of the spectrosocopic work in Kodaikanal was concerned
with the measurements of these minute shifts.

It turns out that the astronomers of the time had no idea of the actual value of pressure in
the various layers of the solar atmosphere. A commonly accepted notion was that the pressure
was several times EarthÕs atmospheric pressure. It was also thought for some time that the
weak lines formed at lower pressures than did the strong lines. Only after the appearance of
Meghnad Saha's work on thermal ionization and its great success in establishing a quantitative
effective temperature scale of stars, were spectroscopists able to calculate the actual pressure
from spectral diagnostics and it turned out that the pressure in the solar atmosphere was only
about a tenth of that in the Earth's atmosphere. The solution to the problem of redshifts came
from an altogether different quarter. Albert Einstein had calculated the gravitational redshift
of spectral lines in his general theory of relativity. As his theory gained credibility, particularly
after the solar eclipse expedition by Eddington in 1919 that gave an accurate measure of the
bending of a light beam in the gravitational field of the Sun, the effect of gravitational redshift
on the spectral lines in the Sun became a major target of study. Towards the end of his stay
at Kodaikanal, Evershed made a comparative study of plates taken in 1914, 1921, 1922 and 1923
and found that for the weaker lines which are supposed to form deep down in the atmosphere,
the results of the redshifts were in agreement with Einstein's theory. However, there was an
unexplained extra redshift detected in limb spectra whose origin was not known.

In January 1911, Michie Smith retired and Evershed became the Director of the Kodaikanal
Observatory. T. Royds joined as the new Assistant Director. In the same year Evershed put
into operation a second spectroheliograph as an auxiliary to the Cambridge instrument utilising
its perfect movement and using a grating and special arrangements for getting photographs
of the solar disk in H$_\alpha$. As red-sensitive plates became available, he started obtaining daily
spectroheliograms in H$_\alpha$ along with the same in the Ca II K line. Several spectacular eruptions
were recorded in some of which the speed of recession of the flying fragments was measured.

\subsection{Prominences}
The study of prominences had continued uninterrupted into the Kodaikanal days. Mary Evershed too
developed an interest in the prominences after moving to Kodaikanal. In 1913, she published
a paper in which she analysed observations of prominences associated with sunspots made
between 1908 and 1910. The number of prominences recorded at Kodaikanal during the years
1904-1914 numbered nearly 60,000 and these, supplemented by the earlier Evershed collection
at Kenley, formed the subject of an exhaustive study and was jointly published by Mary and
John Evershed as a Memoir of the Kodaikanal Observatory in 1917.

In 1913, the Eversheds visited Kashmir on three months leave and found the observing
conditions in the Kashmir Valley excellent for solar work. As Evershed wrote later, ``The Valley
of Kashmir is a level plain containing a river and much wet cultivation of rice. It is 5,000 ft
above sea level and is completely surrounded by high mountains. Under these conditions the
solar definition is extremely good at all times of the day, and unlike most high level stations
it is best near noon and in hot summer weather.'' He established a temporary observatory near
Srinagar in 1915-16 obtaining very high-quality photographs of prominences and sunspots.
This experience of Evershed was later extended to various localities including one on an ocean
liner in tropical waters. The view that the best solar definition is found in low level plains near
the sea or on small islands surrounded by extensive sheets of water grew out of these
experiences.

In 1915, Evershed was elected a Fellow of the Royal Society and he was awarded the Gold
Medal of the Royal Astronomical Society in 1918. At the award ceremony the President of the RAS,
Major P. A. MacMahon, gave a full account of Evershed's work up to that time giving
maximum prominence to the discovery of the radial motion in sunspots. The other contributions
that he highlighted were (i) prominence observations, (ii) spectra of sunspots in general, (iii)
investigations into the reversing layer, (iv) the minute displacements of lines at all parts of
the solar disk, (v) eclipse and cometary observations, and (vi) the invention and perfection
of instruments of observation and measurement. Major MacMahon compared Evershed with
Sir William Huggins saying ``There is much in our medallistÕs career which is a reminder of
the scientific life of William Huggins. They come from the same English neighbourhood, and
began as amateurs of the best kind. They both possess the same kind of scientific aptitude.''

John Evershed retired in 1923 and left Kodaikanal for England. As the Eversheds went down
the ghaut road, a tiger darted across their path, the only time in 16 years in India that they
saw a tiger. They returned to Surrey, where Evershed established his own obsevatory in
Ewhurst near Guildford and continued with his solar observations for well over another thirty
years. He made liquid-filled prisms which revealed great possibilities for high dispersion work.
He obtained Zeeman spectra of sunspots using these prisms as dispersers. He and Mary
continued to take part in solar eclipse expeditions and also attended meetings of the IAU
every three years. Mary died in 1949 at the age of 83 and John Evershed in 1956 at the age
of 92. 

\section{Scientific impact of the Evershed Effect}
As we look back at the spectroscopic detection, through Doppler shifts, of gas motions in
the penumbral photospheric layers of sunspots in 1909 by John Evershed at the Kodaikanal
Solar Observatory, this discovery clearly stands
out as the earliest successful observation of velocity fields due to a complex
magnetohydrodynamic phenomenon in action in an astrophysical setting. The most remarkable
aspect of EvershedÕs work was his ingenuity, given the limited resources and sensitivity
of the instruments that he used, in providing accurate interpretations of his observations that
the gas motions in sunspot penumbrae were indeed radial outflows parallel to the solar
surface. This dominant dynamical phenomenon discovered by Evershed, can be clearly detected
using modern observations with vastly improved spectral and spatial resolution, though additional
components involving vertical and wave motions are also seen in the spectra. Yet, a century later we still do not have a clear physical understanding of this phenomenon.

The immediate scientific impact of Evershed's discovery pertained to another momentous
event in astrophysics, viz. the detection of magnetic field in sunspots by George Ellery Hale
a year earlier in 1908 at the Mt. Wilson Observatory. This was the first ever discovery of a
magnetic field outside the terrestrial environment, and was motivated by the prevailing speculations at that time
that sunspots were giant vortices and that all rotating heavenly bodies harboured magnetic
fields. Hale considered his detection of magnetic fields in sunspots, in turn, as an indication
of swirling flow of ionized gas, which could have generated the magnetic field locally. However,
as Evershed  pointed out in his discovery paper of 1909,  his observations of radial
outflow of gas made the circular flow hypothesis untenable. It was not until the early 1930Õs
that the local generation of magnetic field due to axisymmentric motions was shown to be
impossible by Cowling through theoretical calculations, and in particular, he suggested sunspots
to be composed of  tubes of magnetic flux breaking through the surface.

\begin{figure}  
\begin{center}
\includegraphics[width=\textwidth]{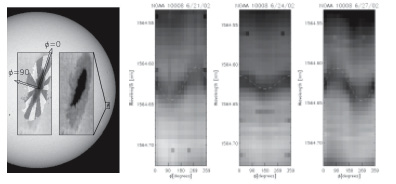}
\caption[]{A modern observation in the CN absorption line at 1546.6 nm of the Evershshed effect depicting the radial outflow of material in a sunspot penumbra (observed by Penn et al. 2003, ApJ, 590, L119)} 
\end{center}
\end{figure}

The two momentous discoveries by Hale and Evershed occurred in an exciting era
during the early 20th century when spectroscopic techniques had begun to be used to study the Sun. An immediate follow-up of research  
on the  Evershed effect was through intense observational investigations, particularly
at the Mt. Wilson Observatory. More observational programmes followed gradually 
world-wide, and in general, observational studies continued over the past
century. However, theoretical attempts to explain the physical causes and understand the
origin of this effect have gone through several inert periods. We are currently in a period
of revived attention mostly fuelled by our increased numerical computational capabilities to
simulate MHD processes.

EvershedÕs initial estimates of flow speeds were of the order of 2 km/s, and he found increasing
speed as distance increases from the spot centre, and no tangential compononets for flow
velocity. His later results, using chromospheric lines, indicated a radial inflow of gas at these
heights; and he measured small tangential components in photospheric velocity, but considered
them unreliable because they were irregular.
Detailed observational programmes started in Mt.Wilson (St.John 1913) and across several
countries in Europe, most notably - at Arcetri in Italy (G. Abetti 1929), at Oxford using the
Oxford Solar Telescope (Kinman 1952), in France and in the Soviet Union in the early 1960Õs,
and at Oslo Solar Observatory during the mid 1960Õs. All these observational studies essentially
confirmed the results of Evershed, added more details on the spatial variation of flow over
the penumbra, established the filamentary structure of the flow, and tentatively identified the
magnetic and flow field associations.
Considerable improvements in spatial and spectral resolutions emerged in the late sixties, and
with that came the identification that Evershed flow is concentrated into dark penumbral
filaments  and where the magnetic field is more horizontal.

Starting from the 1990Õs, high resolution observations, from various groups all over the world,
have served to spatially resolve the flow and magnetic structures which have the following
dominant properties: the flow occurs mainly in the dark penumbral filaments where the field
is more nearly horizontal than the average, the inner penumbral heads of horizontal flow
structures have concentrated but bright upflows, which feed the horizontal flows.

Even though a dominant role for the magnetic field in the thermal and dynamical structures
of a sunspot was envisaged early on by Biermann, Cowling, 
it was the work of Hannes Alfv\'en that was largely responsible for the development of the new
field of magnetohydrodynamics, that provided the conceptual framework for analysing this pheomenon. Theoretical interpretations of the Evershed flow started relatively late
in 1955 when Sweet found an origin in convectively driven motions. Since then, 
various magneto-convective processes have been proposed - they form a large body of research
on sunspot related problems. 

There have been several different theories based on magneto-convective mechanisms to
explain th Evershed flow, most notable ones being: (i) a linear theory of convective rolls in
horizontal magnetic field proposed by Danielson  and its non-linear modification by
Galloway to drive a radial outflow, (ii) BusseÕs theory of three dimensional
convection in inclined magnetic field, where Reynolds stresses drive an Evershed flow, and
(iii) interchange convection  where
a dynamical evolution of thin flux tube produces the Evershed flow. Currently, a new magnetoconvective
picture of Evershed flows has emerged based on 2D and 3D MHD simulations ( discussed in
greater detail in this monograph), where
the convective interactions between an upward hot plume and magnetic field produce all the
necessary ingredients to drive a horizontal Evershed flow as observed. A complete and fully consistent picture of Everhsed flows, with ability to match all the
observed features is yet to be realised, even a 100 years after its discovery. It is likely that
more delicate and detailed picture of Evershed flows would emerge from the very high spatial
and spectral resolution spectropolarimetric observations planned in the near future. It is
equally likely that the fast developing computational resources and algorithms could get
complex enough to reproduce well the observations.

\section{Kodaikanal Observatory: 1923-1960}
Following John Evershed's retirement in 1923, activity in solar physics continued unabated at
Kodaikanal and work progressed along the lines of the early years of the century. The successive
Directors at the Observatory were T. Ryods (1923-1937), A. L. Narayan (1937-1946) and A. K. Das
(1946-1960). The scientific highlights of this era were (a) discovery of oxygen lines in emission in
the chromosphere without the aid of an eclipse, (b) centre to limb variations of hydrogen lines
and their use to study the solar atmosphere, and (c) detailed study of the properties of the
filaments, seen in H$_\alpha$. Significant progress was made on the instrumentation front during this period; the new ionospheric and geomagnetic laboratory was set up in the mid-fifties and a major solar facility, the solar tunnel telescope was commissioned in 1960.

\subsection{Eclipse expeditions}
Eclipse studies constituted an important activity of the Observatory.  
Royds led expeditions to the eclipses in 1929 to Siam, and in 1936 to Japan.
While the totality at Siam was lost to clouds, in Japan, Royds used one of the largest ever
spectrographs to record the spectra and he was the only observer to get any results. Royds
carried out measurement of lines at the extreme limb of the Sun. The 1952 totality in Iraq and the
1955 one in Sri Lanka were again lost to cloudy skies. 

\subsection{The Saha Committee}
A committee appointed by the Government of India, with Meghnad Saha as the Chairman,
examined in 1945 a plan for the post-war development of astronomical research and teaching at the existing
observatories and the universities in the country. One of the main recommendations of this committee was
aimed at improving the facilities for solar observations, especially during the first five year plan.
It was proposed to make available a solar tower telescope, a coronagraph, and a laboratory for
solar terrestrial studies in Kodaikanal. Most of these were implemented by 1960.

\subsection{Ionospheric and geomagnetic laboratory}
Another post-independence development carried out at Kodaikanal was the setting up of a
magnetic and ionospheric laboratory in order to study the reaction of the earthÕs ionosphere and
magnetic field to transient solar events. Two aspects of Kodaikanal greatly stimulated much of
the research carried out subsequently in this area. The first one was the ready and immediate
availability of information on solar happenings observed optically or by radio techniques with
the Kodaikanal telescopes. Few places in the world have available the facility of an ionospheric
laboratory and a solar station in close proximity. The advantages in quick experimentation and
inference have been substantial. The second feature is KodaikanalÕs location very close to the
geomagnetic equator; hence several aspects of the ionosphere, so vital in radio communication
can be studied in Kodaikanal with much advantage.

The laboratory went into operation around 1955, just before the International Geophysical Year.
There have been numerous studies of the properties of the ionosphere over low equatorial
latitudes which have improved our understanding of the phenomena governing radio propagation
and its dependence on the solar radiation characteristics. Interestingly, the laboratory once
recorded the geomagnetic effects of a stellar x-ray source.

\subsection{Solar instrumentation}
A Hale spectrohelioscope for visual observations of the Sun was received as a gift from the
Mount Wilson Observatory during 1933-34, for participation in a systematic watch of the Sun by
international cooperation. The H$_\alpha$\  line in the first order of the grating was used for visual
observations of prominences, dark markings, and for solar flare patrol. A line shifter device was
soon added for visual estimation of the Doppler shifts. This instrument was in regular use for the
next sixty years. Transient activity first noticed with this instrument was followed up with fast
sequences using the main spectroheliographs and also to alert the ionospheric and geomagnetic
laboratory on campus.

A  tunnel telescope by Grubb Parsons, purchased in 1958 was commissioned in 1960. It  consists
of a 60 cm diameter two mirror coelostat, mounted on a 11~m high tower, that directs light via a flat
mirror to a 38 cm aperture, f/90 achromat which forms a 34 cm diameter solar image at the focal
plane. A Littrow-type spectrograph with an inverse dispersion of 9 mm/\AA \  in the fifth order of its
grating, and a spectroheliograph capable of giving pictures in a chosen line became available for
use. These have continued to be, for nearly half a century now, the countryÕs main facility for high
spectral resolution observations of the Sun.

The Kodaikanal Observatory continued its programmes of cooperation with the Greenwich,
Cambridge, Meudon, and the Mt. Wilson observatories, especially with regard to exchange of
observations for missing days. The Observatory was responsible for communicating to the IAU
material for publication of H$_\alpha$ flocculi and dark markings, for most part of the above period. The spectrohelioscope observations were sent regularly as inputs for the quarterly bulletins of the
IAU on solar activity. Beginning 1949 May, the Observatory started sending regular coded
messages on solar activity for the benefit of the Meteorological Department, geophysicists, and
radio specialists in the country and abroad. The Observatory continued to regularly publish the
Kodaikanal Observatory Bulletins and also the half-yearly bulletins of solar statistics.

\subsection{Scientific highlights}
Royds carried out laboratory studies of spectra with a view to investigating the cause of
observed displacements of solar lines with reference to lines in the spectra of terrestrial
sources. Royds and Narayan later observed the centre to limb variations of several
strong solar lines. Royds succeeded in photographing, outside an eclipse, the infra red triplet of oxygen lines at 7771, 7774, and 7775 \AA\  as emission lines in the chromosphere. These lines are
normally observed as absorption lines in the Fraunhofer spectrum.

\begin{figure}  
\begin{center}
\includegraphics[width=0.6\textwidth]{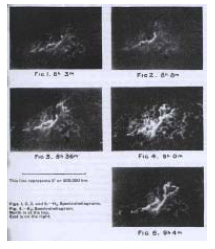}
\caption[]{Royd's flare of Feb. 22, 1926 covering two large sunspot groups is one of the largest ever recorded at Kodaikanal (from IIA archives)} 
\end{center}
\end{figure}

Rao and his team observed H$_\alpha$, H$_\beta$, D, and several other lines near simultaneously during the maximum phase of a solar flare. They argued that the observed line broadenings
were mostly due to Doppler effect and were not assignable either to the Zeeman effect or
to the Stark effect. Narayan and co-workers studied several atoms, ions, and molecules in the laboratory
and looked for their signatures in the solar spectrum. They established the presence of
P$_2$ and studied CN spectral lines in the Sun.

Das and Ramanathan observed the distribution of radiation flux across sunspots in the
continuum and narrow band windows of the H and K lines. They demonstrated that
certain sunspots show the presence of bright rings and that they are especially more
intense in the wavelength of K$_2$. Statistical studies of solar features, such as sunspots, prominences, and filaments and their periodicities, longitudinal and latitudinal distributions, migration patterns, changes in orientations, and rates of occurrences were carried out.

Flare patrol and follow up observations were carried out routinely at the Observatory.
All efforts were made to not miss particularly strong events. In 1926, Royds obtained extensive spectroheliogram coverage of an exceptionally strong
and widespread flare (which was known for long at Kodaikanal as RoydsÕ flare). The
observations were later used for various studies. In 1949, Das and Raja Rao observed a 3+ flare followed by a great magnetic storm
starting 40 hours later but lasting over two days. Peculiarities of the storm were studied
in detail.

\begin{figure}  
\begin{center}
\includegraphics[width=\textwidth]{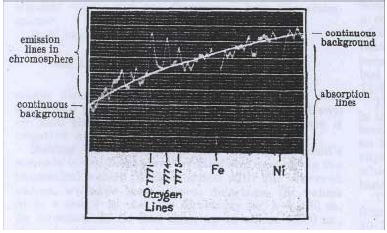}
\caption[]{Microphotometer scan of the spectrum obtained in oxygen emission lines in the chromosphere without the aid of an eclipse. Royds used a straight  slit tangential to the SunÕs limb to record the triplet  (from IIA archives)} 
\end{center}
\end{figure}

In 1953, Das and Sethumadhavan recorded observations of an apparently quiescent
prominence, not visibly associated with any sunspot, grow into a dramatic eruptive
event. Spectra of the dramatic event were recorded and the associated radio noise
bursts were also observed.

\section{Solar physics research at Kodaikanal from 1960 onwards}
M. K. Vainu Bappu arrived in Kodaikanal Observatory in 1960 and by the end of 1962, he put into
operation the newly installed solar instruments, especially the solar tunnel telescope. While solar
observations continued with these facilities, new horizons were opening up in the area of night
time astronomy.

\begin{figure}  
\begin{center}
\includegraphics[width=\textwidth]{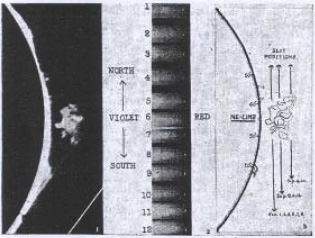}
\caption[]{An apparently quiet prominence suddenly grew to twice its height and erupted. Doppler shifts were measured from the H$_\alpha$. Associated radio noise bursts were also studied (from IIA archives)} 
\end{center}
\end{figure}

Renewed vigor and productivity characterised solar research in Kodaikanal under the leadership
of Bappu. The solar tower telescope was in regular operation in 1962, and by 1965 it
had a spectrograph and a Babcock type magnetograph. There were three spectroheliographs
housed in the building adjacent to the solar tower. The photoheliograph of 20 cm diameter
was also in operation.

\subsection{Velocity fields in he solar atmosphere}
Bhatnagar undertook a detailed study of the Evershed effect based on observations
in Zeeman insensitive lines of Ni I and Fe I. He measured sizeable tangential velocities in the
sunspot penumbrae, with a maximum value of about 0.6 km/sec, and studied in detail the line
asymmetries, which were first observed by Evershed in 1916. Bhatnagar also performed
correlative studies of continuum brightness and equivalent widths and found that darker
regions of the penumbra show larger equivalent widths while an opposite correlation was
found for the brighter regions. This was one of the earliest identification of flow and brightness
associations that have now been well studied by high spectral and spatial resolution
observations.
Studies on solar 5-minute oscillations were undertaken by Sivaraman during 1966 - 1972,
culminating in a Ph.D. thesis. A comprehensive study of various properties of the solar
velocity and intensity oscillations was done, and though it fell short of bringing out the
resolved k-w diagnostic spectrum and thus the resonant-cavity nature of solar interior for
acoustic waves, this work was among the early observational studies of solar oscillations.
Temperature and velocity fields in the temperature-minimum region and the low chromosphere
were studied using photoelectrically determined profiles of molecular lines of CN and C2 by
Nirupama Subrahmanyam in 1965.
Studies of solar rotation, by means of sunspot observations and estimation of their anchor
dpeths in the convection zone, have formed a major research activity in more recent years and
have been carried out by Sivaraman, Gupta and Howard The white light images recorded at Kodaikanal, from early 1900Õs, have been used in a number
of studies including, solar surface and internal rotation, evolution of large scale bipolar
regions, solar diameter, solar activity cycle dependent zonal velocity bands, optical flares,
correlations between rotation of bipolar sunspots and flares.

\subsection{Magnetic fields}
Regular observations of strong magnetic fields in sunspots began in 1963. Measurements were
made by means of the compound quarter-wave-plate technique and using the 6303 A line with
a major objective of evaluating the spatial distribution of longitudinal magnetic field. A
longitudinal magnetograph of the Babcock type was designed and brought into use by J. C.
Bhattacharyya in 1965. It operated at a chopping frequency of 50 cycles/sec and was in use
for the study of weak longitudinal magnetic fields on the solar surface. Weak field in polar
regions and elsewhere were also studied by Bhattacharya.

\subsection{Chromospheric features and dynamics}
Exhaustive studies of Ca II K line spectra were carried out during 1969-1976 by Bappu and
Sivaraman. The analyses and results included: role of Ca II K line as a reliable diagnostic of
chromospheric activity, the solar cycle variation of Ca II K line emission profiles of integrated
sunlight, association between photospheric magnetic structures and Ca II K structures and
their use in inferring the morphological evolution of photospheric structures, and the implications
of the above for other stars. Variation of the luminosity of the Sun as star in Ca II K line was
studied by  Sivaraman, Singh,  Bagare, and  Gupta in the mid-eighties.
During this same period, Jagdev Singh and co-workers carried out a detailed study of active region
contributions to Ca II K line luminosity variations over a solar cycle. Intensity fluctuations
in the H$_\alpha$\ line due to chromospheric mottling were studied by  Bappu in 1964. Singh
and Bappu studied the supergranular network size and its variation with solar cycle.
Bappu also carried out detailed observations of prominences and calcium flocculus in the early
sixties. Sivaraman and Makarov observed filament migrations, and poleward migration
of magnetic structures, meridional motions, and polarity reversals at the poles.

\subsection{Coronal studies}
Bappu, Bhattacharyya, and Sivaraman observed  H$_\alpha$ \  and Ca II K lines during the solar eclipse of 1970. Important coronal studies, most notably on coronal wave dynamics, were carried out over several eclipse expeditions starting from 1983 to date. The recent eclipse expeditions have achieved high spatial resolution narrow band photometry of coronal structures to investigate the the nature of waves. Studies of intensity oscillation in the coronal green
line and red emission line have also been carried out.

\subsection{Solar terrestrial studies}
Several studies relating to the ionosphere over Kodaikanal and the electrojet have been carried
out in the early sixties. Notable among them are the variation over a solar cycle of ionic
densities at different levels of the equatorial ionosphere, and the spatial, seasonal, and solar
cycle variations in the lunar semi-diurnal oscillations in the ionospheric F-region. An analysis
of magnetic crochets associated with relativistic flares was done in 1964 and the amplitude
characteristics of geomagnetic sudden commencements with energetic proton events were
studied.

\subsection{Current observational facilities and programmes}
The Ca II K, H$_\alpha$\  images and spectroheliograms are continuing to be obtained at Kodaikanal for studies of the chromospheric network, solar cycle variations of the background flux in the K line, and synoptic observations of solar activity. Digitisation of Kodaikanal data is one major new project: a new digitiser using better CCD camera and uniform light source have been fabricated to digitise the solar images with higher resolution (spatial resolution 1 arc sec, limiting resolution of the images) and with very high photometric accuracy.

A new spectropolarimeter, indigenously designed and built, has been installed and being used
at Kodaikanal for the study of active regions on the Sun. A new system to obtain K-line images
of the Sun using a K-line filter has been installed. These K-filtergrams are recorded onto a
1K $\times$ 1K CCD. The system has been operational since 1996-97.

The main research areas include:
\begin{itemize}
\item Oscillation in the chromospheric network;
\item Solar cycle variations and synoptic observations of solar activity;
\item Dynamics of the solar corona and coronal holes;
\item Sunspots and local helioseismology;
\item Solar interior;
\item Coronal mass ejections.
\end{itemize}

\section{Future programmes}
\subsection{National Large Solar Telescope (NLST)}
The National Large Solar Telescope NLST will be a state-of-the-art 2-m class telescope for carrying out high-resolution studies of the solar atmosphere. Sites in the Himalayan region at altitudes greater than 4000-m that have extremely low water vapor content and are unaffected by monsoons are under evaluation. This project is led by the Indian Institute of Astrophysics and has national and international partners. Its geographical location will fill the longitudinal gap between Japan and Europe and is expected to be the largest solar telescope with an aperture larger than 1.5 m till the 4-m class Advanced Technology Solar Telescope  (ATST) and the European Solar Telescope (EST) come into operation.

NLST is an on-axis alt-azimuth Gregorian multi-purpose open telescope with the provision of carrying out night time stellar observations using a spectrograph at the Nasmyth focus. The telescope utilizes an innovative design with low number of reflections to achieve a high throughput and low polarization. High order adaptive optics is integrated into the design that works with a modest FriedÕs parameter of 7-cm to give diffraction limited performance. The telescope will be equipped with a suite of post-focus instruments including a high-resolution spectrograph and a polarimeter. A small (20cm) auxiliary telescope will provide full disk images.

The detailed concept design of the telescope is presently being finalized. First light is expected in 2013.

\subsection{Space Coronagraph}
A visible emission line coronagraph that uses an innovative design to simultaneously obtain images of the solar corona in the Fe XIV green emission line at 530.3nm and the Fe X red line at 637.4 nm is under development. The mission is capable of taking images in the visible wavelength range covering the coronal region between 1.05 to 3 solar radii with a frequency of 4 Hz using an efficient detector. High cadence observations in the inner corona are important to understand the rapidly varying dynamics of the corona as well as to study the origin and acceleration of CMEs. There are currently no such payloads planned for the near future. 

This 20-cm space coronagraph, that will be executed under the leadership of the Indian Institute of Astrophysics, is planned for launch in 2012. It will obtain simultaneous images of the solar corona in the green and red emission lines simultaneously with a field of view between 1.05 -1.60 solar radii to: (a) study the dynamics of coronal structures; (b) map the linear polarization of the inner corona; and (c) monitor the development of CMEÕs in the inner corona by taking coronal images with high cadence up to 3 solar radii.

The large telemetry capability of the dedicated mission will permit a monitoring of CMEs for about 18 hours a day. This project with several national partners, has been accepted in principle by the Indian Space Research Organization.

\end{document}